\documentclass[tighten,twocolumn]{aastex63}

\usepackage{graphicx}
\usepackage{ulem}
\usepackage{amsmath}
\usepackage{wrapfig}
\usepackage[thinlines]{easytable}
\usepackage{tikz}
\usepackage{hyperref}
\usepackage{tabularx}
\usepackage{url}
\usepackage{lineno}
\usepackage{svg}

\graphicspath{{./}{figures/}}

\begin{document}

\title{Red vs. Blue: How metallicity shapes black hole dynamics and mergers in dense star clusters}

\correspondingauthor{Saloni Agrawal}
\email{saagrawal@ucsd.edu}

\author[0009-0007-6748-4627]{Saloni Agrawal}
\affiliation{Department of Astronomy \& Astrophysics, University of California, San Diego, La Jolla, CA, USA}
\author[0000-0002-4086-3180]{Kyle Kremer}
\affiliation{Department of Astronomy \& Astrophysics, University of California, San Diego, La Jolla, CA, USA}
\author[0000-0002-0147-0835]{Michael Zevin} 
\affiliation{Adler Planetarium, 1300 South DuSable Lake Shore Drive, Chicago, IL 60605, USA}
\affiliation{Center for Interdisciplinary Exploration \& Research in Astrophysics (CIERA), Northwestern University, Evanston, IL 60201, USA}
\author[0000-0002-1144-6708]{Cailin Plunkett}
\affiliation{LIGO Laboratory, Massachusetts Institute of Technology, Cambridge, MA 02139, USA}
\affiliation{Kavli Institute for Astrophysics and Space Research, Massachusetts Institute of Technology, Cambridge, MA 02139, USA}
\affiliation{Department of Physics, Massachusetts Institute of Technology, Cambridge, MA 02139, USA}
\author[0000-0002-0933-6438]{Elena Gonz\'{a}lez Prieto}
\affiliation{Department of Physics \& Astronomy, Northwestern University, Evanston, IL 60208, USA}
\affiliation{Center for Interdisciplinary Exploration \& Research in Astrophysics (CIERA), Northwestern University, Evanston, IL 60201, USA}
\author[0000-0003-4412-2176]{Fulya K{\i}ro\u{g}lu}
\affiliation{Center for Interdisciplinary Exploration \& Research in Astrophysics (CIERA), Northwestern University, Evanston, IL 60201, USA}
\author[0000-0003-3987-3776]{Christopher E.\ O'Connor}
\affiliation{Center for Interdisciplinary Exploration \& Research in Astrophysics (CIERA), Northwestern University, Evanston, IL 60201, USA}
\author[0000-0002-7132-418X]{Frederic A.\ Rasio}
\affiliation{Department of Physics \& Astronomy, Northwestern University, Evanston, IL 60208, USA}
\affiliation{Center for Interdisciplinary Exploration \& Research in Astrophysics (CIERA), Northwestern University, Evanston, IL 60201, USA}
\author[0000-0001-9582-881X]{Claire S.\ Ye}
\affiliation{Canadian Institute for Theoretical Astrophysics, University of Toronto, 60 St.\ George St., Toronto, ON M5S 3H8, Canada}

\begin{abstract}
Dense star clusters are a well-established environment for the formation of gravitational wave sources through dynamical interactions. Recent LIGO--Virgo--KAGRA (LVK) events such as GW241011 and GW241110 provide some of the best evidence yet for a dynamical origin.
However, their relatively low component masses are in tension with predictions from low-metallicity globular cluster models (which typically produce more massive black holes), hinting that these events may have originated in higher-metallicity environments.
Here we present a new set of Monte Carlo star cluster simulations with refined coverage in metallicity, focusing specifically on clusters with [Fe/H] $\geq-1$, similar to the ``red'' globular cluster subpopulation observed in most galaxies. We show that metallicity has a significant effect on the mass function of black holes and black hole mergers, the total number of black hole mergers per cluster, black hole retention from natal kicks, the mass segregation time for black-hole-driven cluster dynamics, and the merger delay time distribution.
We also show that high-metallicity cluster models produce low-mass hierarchical mergers consistent with the mass ratios and component masses of GW241011 and GW241110, motivating the importance of high-metallicity clusters in the astrophysical interpretation of future LVK catalogs.
\vspace{1cm}
\end{abstract}


\section{Introduction}
\label{sec:intro}
It is now widely accepted that the dense centers of globular clusters host significant populations of black holes formed via the collapse of massive stars \citep[e.g.,][]{Strader2012,Giesers2018,Giesers2019,Askar2018,Kremer2019,Weatherford2020,Vitral2022,Dickson2024,Kremer2025}. With their large masses and high central densities, globular clusters provide an efficient environment for frequent dynamical encounters to form and harden binary black holes into observable gravitational-wave sources detectable by the LIGO--Virgo--KAGRA (LVK) collaboration \citep[e.g.,][]{PortegiesZwart2000,Rodriguez2016,Askar2017,Samsing2018,Kremer2020_catalog,AntoniniGieles2020,Ye2026,LIGO2016,LIGO2025}. 

A characteristic signature of this dynamical channel is the presence of hierarchical mergers, in which the remnants of first generation (1G) black hole mergers (i.e., the black holes that formed directly from stellar collapse) are retained within their host clusters, allowing second generation (2G) and higher-generation black holes to participate in subsequent mergers 
\citep[e.g.,][]{Rodriguez2019,Fragione2020,Gerosa2021,Mai2026}. Hierarchical mergers have been invoked to explain a number of features in the LVK data \citep[e.g.,][]{Fishbach2017,Kimball2021}, such as events like GW190521 and GW231123, which reside in the mass range expected to be subject to pair instability supernovae \citep[e.g.,][]{GW190521_2020,GW231123_2025}.

The recent gravitational wave events GW241011 and GW241110, which exhibit high primary spins consistent with expectations for a binary black hole merger remnant, non-negligible spin-orbit misalignment consistent with an underlying isotropic spin distribution, and asymmetric mass ratios consistent with expectations for a merger between a 2G and a 1G black hole, are perhaps the best evidence yet for hierarchical merger origins \citep{LIGO2025GW241011}. Interestingly, their component masses---roughly $20\,M_{\odot}+6\,M_{\odot}$ and $17\,M_{\odot}+7.7\,M_{\odot}$, respectively---place them in a mass regime lower than typically associated with dynamical assembly in low-metallicity globular cluster environments \citep[e.g.,][]{Rodriguez2016,Chatterjee2017}. 

A natural way to produce hierarchical mergers at these relatively low masses is through higher-metallicity star clusters \citep{LIGO2025GW241011,Ye2026}. As stellar metallicity increases, line-driven winds from massive stars are expected to become stronger, increasing mass loss prior to core collapse \citep[e.g.,][]{Vink2001,VinkDeKoter2005,Mokiem2007}. Because these winds reduce the pre-supernova stellar and core masses, stellar-evolution models generally predict that black hole masses (both the average and the maximum black hole mass) are lower at higher metallicity \citep[e.g.,][]{Heger2003,EldridgeVink2006,Belczynski2010,Spera2015,SperaMapelli2017}. Although winds remain the dominant effect, the complete physics governing the final remnant-mass distribution is multifaceted, with many other effects contributing, such as binary mass transfer and stripping, rotation and internal mixing, core-collapse/fallback physics, and (pulsational) pair-instability mass loss \citep[e.g.,][]{Fryer2012,Sukhbold2016,Vartanyan2018,BurrowsVartanyan2021,Farmer2019,MarchantBodensteiner2024}. However, the general expectation is that metal-poor populations  ([Fe/H] $\lesssim-1$) produce a larger fraction of relatively massive black holes in the $20$--$40\,M_{\odot}$ range, while near-solar-metallicity populations produce predominantly lower-mass remnants in the $5$--$15\,M_{\odot}$ range. Current observational evidence for a direct black hole mass--metallicity relation remains sparse and selection-biased; however, this qualitative trend is consistent with the mass distribution of near-solar Galactic black holes in X-ray binaries \citep[e.g.,][]{Farr2011,CorralSantana2016}, the detached Gaia BH1 and BH2 systems \citep{ElBadry2023a,ElBadry2023b}, and the metal-poor, high-mass Gaia BH3 system \citep{Gaia2024}.

\begin{figure}[ht!]
\centering
\includegraphics[width=\linewidth,trim = 0 0 0 0,clip]{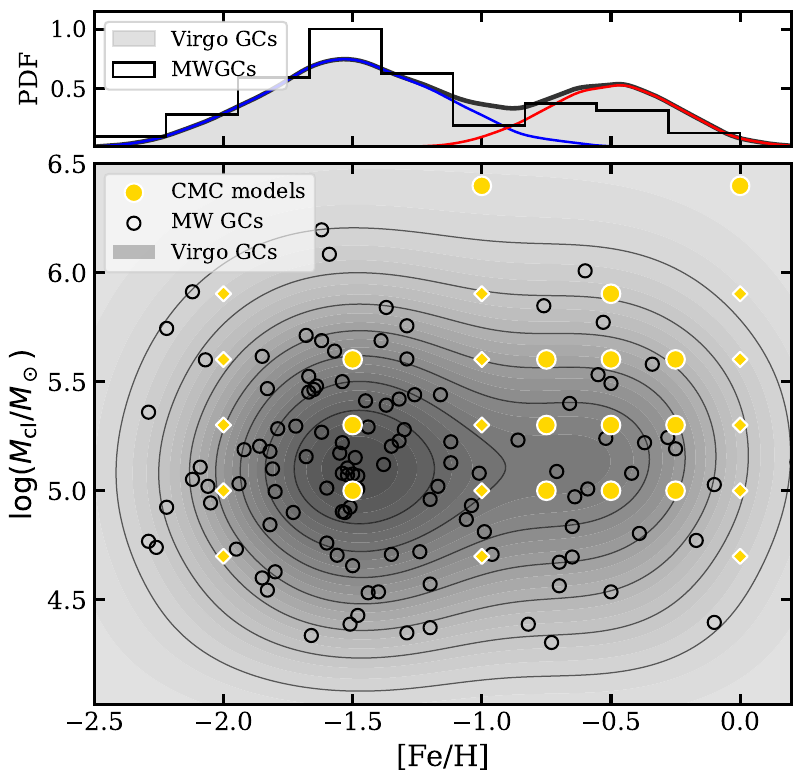}
\caption{Metallicity and mass distribution of observed and simulated globular clusters. The upper panel shows the probability density (PDF) for metallicity values for the full globular cluster population in Virgo (gray), and separated into metal-rich (red) and metal-poor (blue) clusters \citep[e.g.,][]{Peng2006,Harris2009,Strader2011}. We also show in black the metallicity distribution for Galactic globular clusters \citep{Harris1996}. The lower panel compares Virgo (gray contours), Milky Way (black circles), and \texttt{CMC} globular cluster models (yellow markers). Yellow diamonds represent \texttt{CMC Cluster Catalog} models described in \citet{Kremer2020_catalog}, and circles represent the new models computed in this study to refine our coverage in metallicity.}
\label{fig:GC_FeH}
\end{figure}

Although prototypical old, halo globular clusters are metal poor, globular cluster populations in massive galaxies commonly contain both metal-poor and metal-rich clusters, often appearing as bimodal color or metallicity distributions \citep[e.g.,][]{BrodieStrader2006,Peng2006,Usher2012}. Roughly $25\%$ of Milky Way globular clusters have [Fe/H] $>-0.75$ \citep{Harris1996}, including some of the densest and most massive clusters like Terzan~5 with $[\rm{Fe/H}]=-0.23$ \citep{Lanzoni2010_Terzan5}. More massive galaxies that are enriched more quickly contain even more high-metallicity systems, often featuring a 2:1 ratio of globular clusters populating the ``blue'' and ``red'' peaks near [Fe/H] $\approx -1.5$ ($Z \approx 0.03Z_{\odot}$) and [Fe/H] $\approx -0.5$ ($Z \approx 0.3Z_{\odot}$), respectively \citep[e.g.,][]{Cohen1998,Peng2006,Harris2009,Strader2011}. In the most massive elliptical galaxies, the location of both peaks shifts to slightly higher values; for example, recent work shows M87 has blue and red peaks near [Fe/H] values of roughly $-1.2$ and $-0.2$, and extending up to supersolar \citep{Villaume2019}. Furthermore, young massive clusters born relatively recently \citep[for a review, see][]{PortegiesZwart2010} can naturally provide high-metallicity dynamical environments conducive to binary black hole mergers. For example, \citet{Chatterjee2017} performed a set of \texttt{CMC} cluster models across a range of metallicities to show that near-solar metallicity clusters are ideal sites for formation of low-mass LVK events like GW151226 \citep{GW151226}, the second confirmed LVK event detected. Subsequent analyses have confirmed this initial prediction, while also comparing to the latest LVK catalogs \citep[e.g.,][]{Banerjee2017,DiCarlo2019,Rodriguez2023,Ye2026}. Thus, consideration of a broad range of metallicities is crucial for understanding the complete landscape of black hole mergers in dense environments.

In Figure~\ref{fig:GC_FeH}, we show cluster mass and metallicity distributions representative of a sample of local Universe globular clusters.
Milky Way cluster properties (open circles) are taken from \citealt{Harris1996}.
Virgo cluster masses (gray contours) are obtained from the ACS Virgo Cluster Survey \citep{Jordan2009} and we approximate the expected metallicity distribution using a bi-modal Gaussian (shown in the top panel of the figure), with mean and full-width half-maximum values $[\mu,\sigma]=[-1.5,0.35],\,[-0.5,0.25]$ for the blue and red populations, respectively. The locations of these peaks are a reasonable proxy for most extragalactic globular cluster populations \citep[e.g.,][]{BrodieStrader2006}. 
Motivated by the observational peaks near [Fe/H] = $-1.5$ and $-0.5$, we present a new suite of dense star cluster simulations featuring refined coverage across all metallicities. This suite is intended as a supplement to the \texttt{CMC Cluster Catalog} \citep{Kremer2020_catalog}, which originally covered three metallicity values, [Fe/H]$=[-2,-1,0]$, that coarsely span the range of Galactic globular clusters. In the lower panel of Figure~\ref{fig:GC_FeH}, yellow diamonds indicate models from the original \texttt{CMC Cluster Catalog}, while yellow circles indicate the new models introduced in this work. Note this panel shows only a two-dimensional projection of the broader model grid onto initial cluster mass and metallicity (see Section~\ref{sec:methods}).

The rest of the paper is organized as follows. In Section~\ref{sec:methods}, we describe the \texttt{Cluster Monte Carlo} code (\texttt{CMC}) and \texttt{COSMIC} simulations used, our grid of initial conditions, and the dynamical merger channels we identify in the models. In Section~\ref{sec:bhmass}, we characterize how metallicity shapes the black hole mass function under our stellar evolution assumptions, and host cluster properties like core radius and black hole retention. In Section~\ref{sec:bhmergers}, we analyze how metallicity changes the total binary black hole merger yield, merger efficiency, and the merger masses. Finally, in Section~\ref{sec:comparison}, we compare our high metallicity merger populations directly to GW241011 and GW241110. We summarize our conclusions and discuss future work in Section~\ref{sec:discussion}. Following the precedent of the original \texttt{CMC Cluster Catalog}, we publicly release all cluster data and binary black hole merger data from these new models, available for download on \href{https://doi.org/10.5281/zenodo.20739155}{Zenodo}.

\section{Modeling black hole mergers}
\label{sec:methods}

\subsection{Globular clusters with \texttt{CMC}}

To model the long-term evolution of dense star clusters, we use the \texttt{CMC} code \citep[][and references therein]{Rodriguez2022}.
We follow the basic structure (and initial conditions) of the \texttt{CMC Cluster Catalog} \citep{Kremer2020_catalog}.
We emphasize refined metallicity coverage, initializing with [Fe/H] = $[-1.5, -0.75, -0.5, -0.25]$ over a subset of $N,r_v$ values including $N=[4,8,16,32]\times10^5$ and $r_v=[0.5,1,2]\,$pc. Unless otherwise specified, we fix $R_{\rm gc}=8\,$kpc for all new simulations performed here. We also include two additional massive models not included in the initial \texttt{CMC Catalog} with $N=10^7$, $r_v=2\,$pc, $R_{\rm gc}=20\,$kpc, and [Fe/H]$=-1$ and $0$ \citep[for further description, see][]{Mai2026}. All other initial conditions (e.g., the \citet{Kroupa2001} initial mass function, King profile parameters, initial binary fraction of $5\%$, primordial binary properties, stellar/binary star evolution assumptions) follow identically the assumptions in \citet{Kremer2020_catalog}. This adds 28 new simulations to the existing \texttt{CMC Cluster Catalog} grid. Our full list of new simulations, including initial cluster properties and black hole merger data, is shown in Table~\ref{table:1}.

As discussed in previous studies \citep[e.g.,][]{Rodriguez2016,Samsing2018,Zevin2019,Rodriguez2019,Samsing2020,Kremer2026_review}, black hole mergers occur through six distinct dynamical channels in globular clusters which we identify in our \texttt{CMC} models. These distinct channels fall into three broad categories: 
\begin{enumerate}
    \item \textit{In-cluster mergers} can occur in one of two ways. First, a black hole binary can harden due to dynamical encounters and eventually merge due to gravitational wave energy dissipation. We label such mergers as ``in-cluster 2 body'' in Table 1; they do not include black hole binaries that originate from primordial stellar binaries. However, for completeness and for the sake of comparison to \citet{OConnor2026}, we consider ``primordial binaries'' as a separate channel in which a primordial stellar binary within a cluster forms a merging black hole pair via quasi-isolated binary processes. 
    \item Black hole binaries can also form via \textit{gravitational-wave capture} during single-single, binary-single, and binary-binary encounters, which we denote as ``2-body capture'',``3-body capture'', and ``4-body capture.'' 
    \item Lastly, \textit{ejected mergers} refer to binary black holes that are hardened inside the cluster, ejected via dynamical recoil, and merge outside their original host at later times following gravitational-wave inspiral. 
\end{enumerate}

Alongside the formation channel, we also track the merger history of each black hole: First-generation (1G) black holes form directly from stellar collapse, while higher-generation black holes (2G and 3G) are merger remnants. We label binary mergers by the generations of their components: for example, 1G+1G, 2G+1G, 2G+2G, and so on.

\subsection{Isolated binaries with \texttt{COSMIC}}

Single-star and binary evolution is performed using \texttt{COSMIC}~\citep{Breivik2020}, which is integrated into \texttt{CMC}. 
We also perform isolated \texttt{COSMIC} runs with consistent settings to \citealt{Kremer2020_catalog} at the modeled cluster metallicities as a means for comparison to the \texttt{CMC} models (see Section~\ref{sec:bhmergers}). For the \texttt{COSMIC} runs, we consider black holes that were in binaries but merged within a Hubble time and black holes that are still in binaries after a Hubble time. 
Although the latter of these black hole populations did not merge, they help provide a basis for comparison with the black hole mergers in \texttt{CMC}, since dynamical processes can cause such systems to merge. 
Physical prescriptions and choices for various parameterizations in the \texttt{COSMIC} models are summarized in Appendix~\ref{appB}.

\begin{figure}[t!]
\centering
\includegraphics[width=\linewidth,trim = 70 110 180 110,clip]{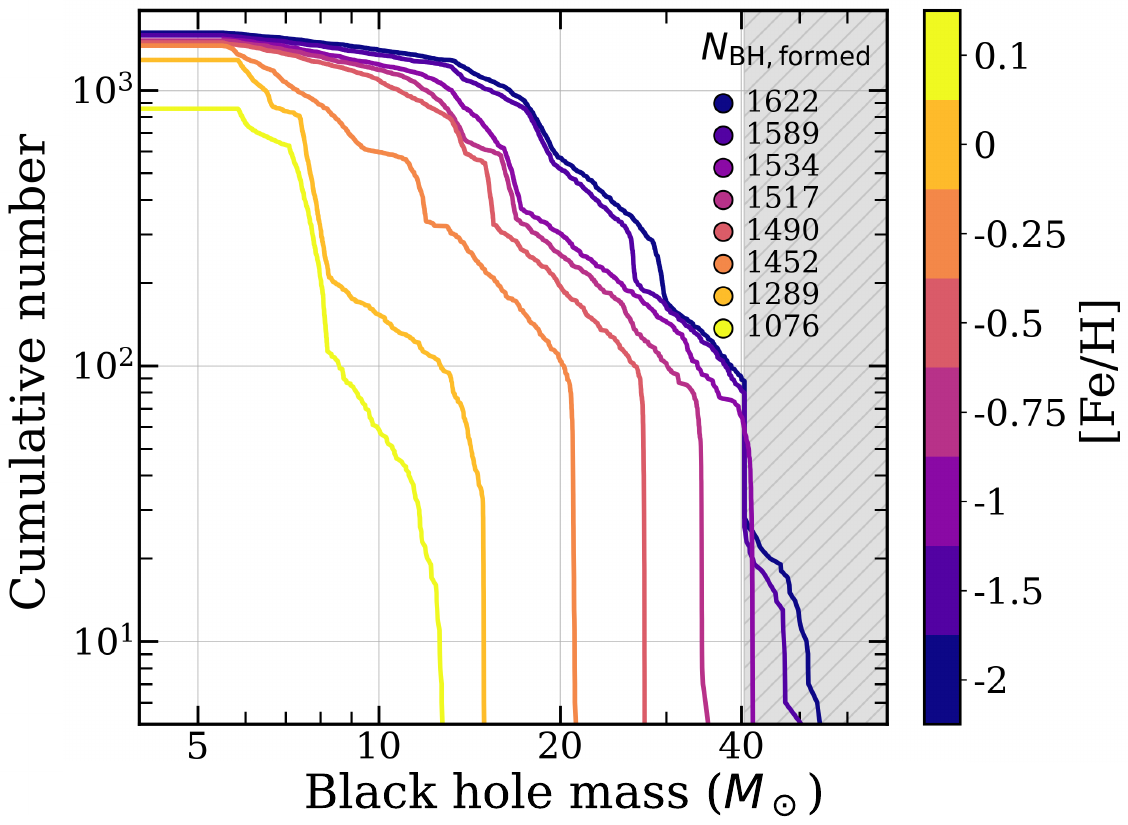}
\caption{Black hole mass function (shown as cumulative distribution) for all black holes formed via stellar collapse in our \texttt{CMC} models with fixed initial conditions, \(N=8\times 10^5\), \(r_v=1\,\mathrm{pc}\), and \(R_{\rm gc}=8\,\mathrm{kpc}\). The eight colors span our full range in metallicity: \(-2 \leq [\mathrm{Fe/H}] \leq 0.1\). The gray shaded region denotes the boundary for the pair-instability mass gap adopted in our models. The mass functions shift toward lower masses with increasing metallicity primarily through our metallicity-dependent stellar-wind prescriptions \citep{Vink2001}. We also list the total number of black holes in each model, which decreases as metallicity increases.}
\label{fig:bhmass}
\end{figure}

\section{Results}

\subsection{Effects of metallicity on black hole masses and cluster properties }
\label{sec:bhmass}

We begin by examining how metallicity affects the black hole population and the dynamical evolution of the cluster. To simplify the comparison across metallicities, we focus on a subset of 8 models --- numbers 51, 158, 55, 159, 160, 161, 59, and 174 [see Table~\ref{table:1}] --- spanning $[{\rm Fe/H}] = [-2,\,-1.5,\,-1,\,-0.75,\,-0.5,\,-0.25,\,0,\,0.1]$; these include the  $[{\rm Fe/H}] = [-2,\,-1,\,0]$ values from the original \texttt{CMC Cluster Catalog}. These models share the same initial conditions:  $N = 8\times10^5$, $r_v = 1\,$pc, and $R_{\rm gc}=8\,$kpc. This subset isolates metallicity as the sole variable driving changes in black hole formation, retention, and merger times.  

Figure~\ref{fig:bhmass} shows the complementary cumulative black hole mass distribution for all black holes formed via stellar collapse across these 8 models (we do not include black holes formed via black hole mergers in this plot) along with the number of black holes formed in each model. The gray hatched area above $40.5\,M_\odot$ indicates our adopted pair-instability mass gap ($45\,M_\odot$ pair instability limit, minus $4.5\,M_\odot$ from neutrino mass loss); black holes within this gray region are formed from stellar collisions at early times \citep[e.g.,][]{Kremer2020_collisions,Gonzalez2021}. As shown, the mass distribution shifts monotonically towards higher masses with decreasing metallicity; the median black hole mass increases from $7.3 M_{\odot}$ at  $[{\rm Fe/H}] = 0$ to $18.1 M_{\odot}$ at  $[{\rm Fe/H}] = -2$, with median values of $M_{\rm BH}/M_{\odot}= [18.1,\,\,  17.9,\,\,  14.8,\,\, 13.5,\,\,  13.3,\,\,  8.8,\,\,  7.5,\,\,7.3]$ at $[{\rm Fe/H}] = [-2,\,-1.5,\,-1,\,-0.75,\,-0.5,\,-0.25,\,0,\,0.1]$ respectively. This behavior primarily reflects our prescription for metallicity-dependent stellar winds (see Section~\ref{sec:intro}): at higher metallicity, stronger line-driven winds reduce the final pre-supernova mass and black hole mass. In addition to reducing the remnant mass, increasing metallicity also decreases the total number of black holes formed in our models (shown as the listed numbers in the figure). This is because at higher metallicity, a larger fraction of massive-star progenitors fall below the threshold for black hole formation and instead leave behind neutron stars.

We note that the precise median masses and high-mass cutoffs shown in Figure~\ref{fig:bhmass} are heavily model-dependent. The mapping between zero-age main sequence (ZAMS) mass, metallicity, and final black hole mass remains highly uncertain and depends on adopted prescriptions for stellar winds, remnant formation, pair-instability physics, natal kicks, and binary evolution. \citep[e.g.,][]{Heger2003,Belczynski2010,Spera2015,Spera2019,Farmer2019,MarchantBodensteiner2024}. Recent population-synthesis calculations, including POSYDON models based on MESA \citep[Modules for Experiments in Stellar Astrophysics;][]{Paxton2011} binary-evolution grid and SEVN-based studies, have shown that variations in the assumed stellar and binary physics impact the predicted ZAMS to remnant mass relation and alter the resulting compact object mass spectrum \citep{GiacobboMapelli2018,Fragos2023,Andrews2025}. Similarly, modern core-collapse calculations demonstrate that compact remnant masses are not a simple monotonic function of the initial stellar mass, but depend on the detailed pre-supernova core structure and explosion/fallback physics \citep{Sukhbold2016,VartanyanBurrows2023}. Thus, while our results recover the expected trend of decreasing black hole mass with increasing metallicity, we caution against over-interpreting the granular features of the various curves in Figure~\ref{fig:bhmass}; we quote precise median masses and high-mass cutoffs for completeness, but these should be interpreted in the context of these theoretical uncertainties.

\begin{figure}
\centering
\includegraphics[width=0.9\textwidth,trim = 220 30 15 45,clip]{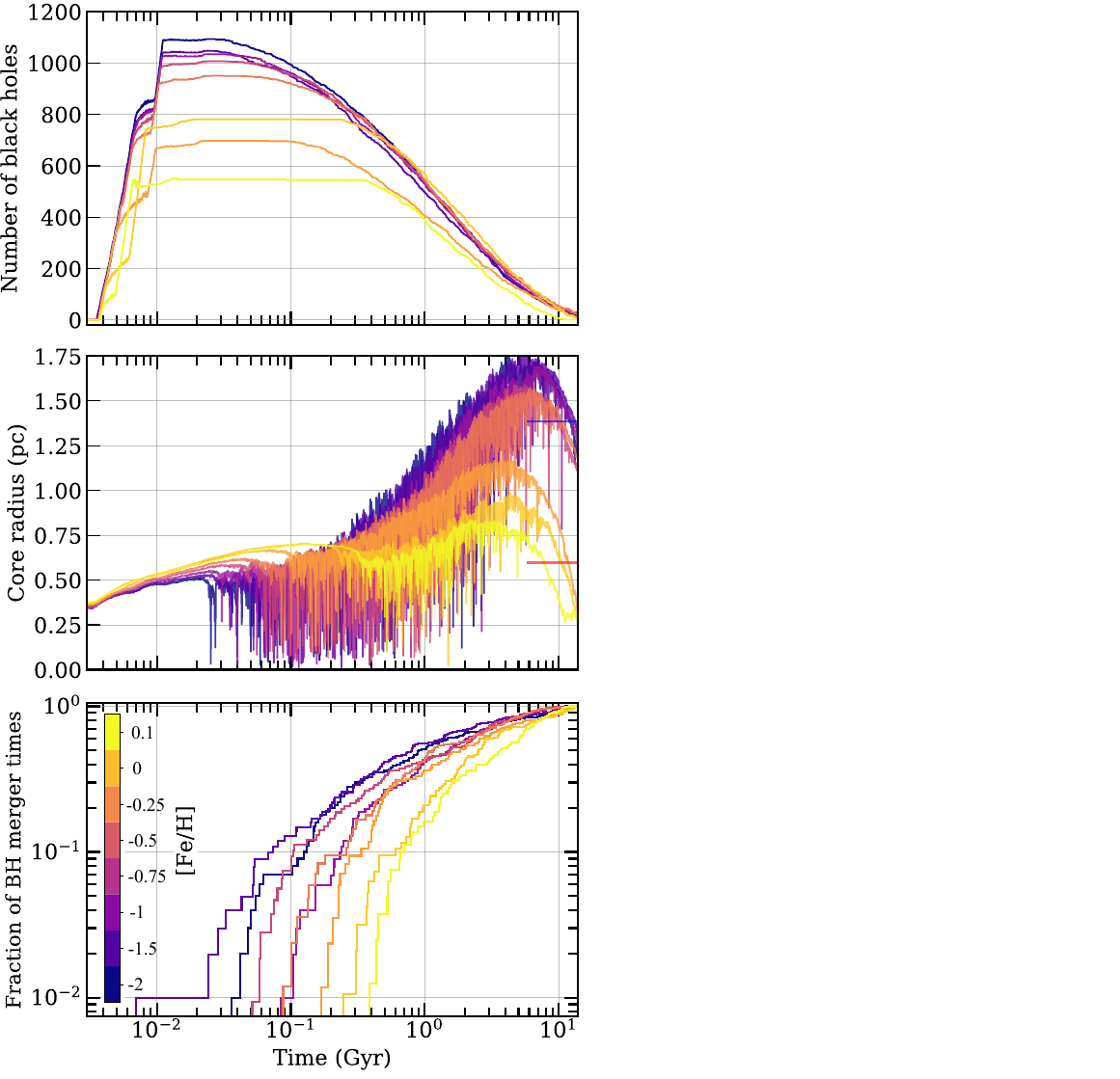}
\caption{Evolution of various cluster properties with time for the same eight \texttt{CMC} models shown in Figure~\ref{fig:bhmass}, covering our full range in metallicity. \textit{Top panel:} number of black holes retained in the host cluster versus time. \textit{Middle panel:} core radius versus time. Notice that lower-metallicity models generally feature more rapid mass segregation at early times and enhanced core expansion at late times, both due to larger black hole masses. They also undergo core collapse at later times due to a higher population of retained black holes. The red (blue) ticks mark the average core radius for Galactic globular clusters at high (low) metallicity. \textit{Bottom panel:} cumulative fraction of black hole merger times. Lower-metallicity models feature shorter delay times, again a result of their relatively large black hole masses.}
\label{fig:Nbh}
\end{figure}

Figure~\ref{fig:Nbh} shows the time evolution of the black hole populations and its influence on the host cluster structure and properties.  We consider the same eight models shown in Figure~\ref{fig:bhmass} (varying [Fe/H], fixed $N=8\times10^5$ and $r_v=1\,$pc). The top panel shows the number of retained black holes as a function of time. The black hole population rises rapidly during the first several Myr as massive stars evolve off the main sequence and collapse. At $t\gtrsim 100\,\rm{Myr}$, the number of retained black holes starts to decrease as dynamical interactions, binary hardening, mergers, and ejections gradually deplete the black hole population. As discussed in Figure~\ref{fig:bhmass}, higher-metallicity models form fewer black holes overall. The top panel further shows that the total number of black holes retained also decreases steadily with increasing metallicity. This is because higher metallicity clusters produce lower-mass remnants that receive larger natal kicks under our fallback prescriptions \citep{Fryer2012}, making them more likely to be ejected at birth.

\begin{figure*}
\centering
\includegraphics[width=1\linewidth,trim = 0 230 8 143,clip]{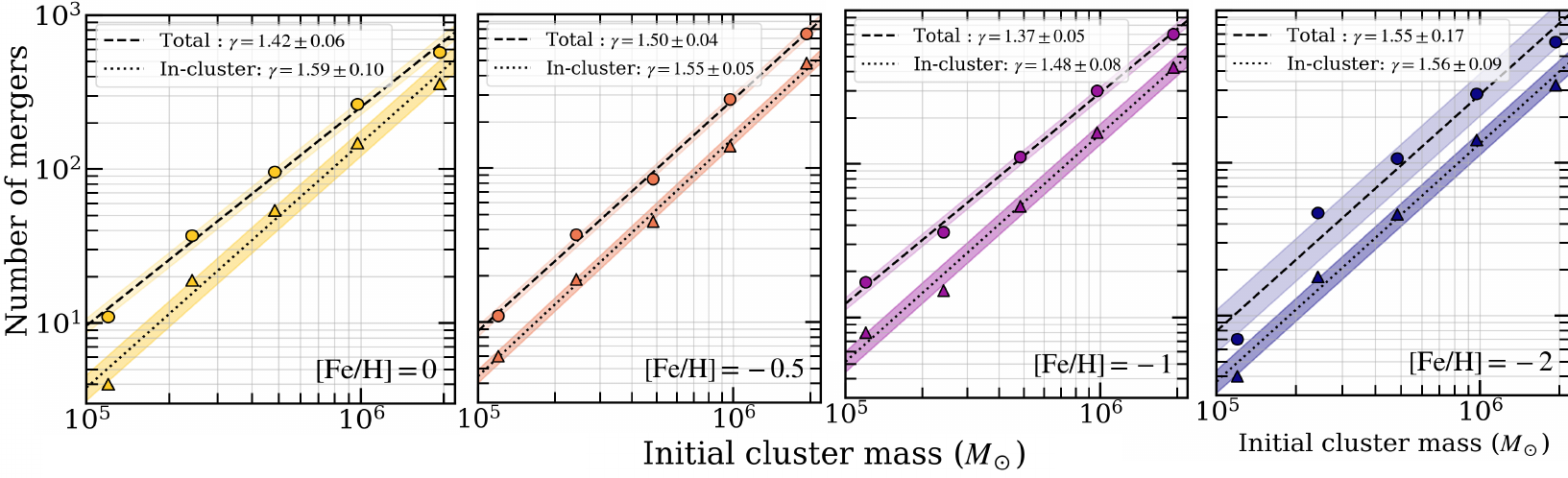}
\caption{Number of total (circles) and in-cluster (triangles) black hole mergers as a function of initial cluster mass across metallicity, showing power-law fits and intrinsic scatter ($1\sigma$) for [Fe/H]$= [0, -0.5, -1, \rm{and}\,-2]$. Although the overall normalization changes (see also Figure~\ref{fig:merg_vs_met}), we see no significant change in slope with metallicity. Here, $\gamma$ denotes the slope of the power-law fit shown in the legend.}
\label{fig:metfit}
\end{figure*}

 The middle panel shows the time evolution of the core radius, $r_c$, reflecting the cluster’s response to early stellar mass loss, black hole mass segregation, black hole binary burning, and eventual core collapse. Here we define core radius using the density-weighted value of \citet{CasertanoHut1985}. This definition is sensitive to transient density oscillations arising from deep collapse of small numbers of black holes in the cluster's innermost regions \citep[e.g.,][]{Morscher2015}; the onset of these sharp density spikes is a visual indicator of the time black hole dynamical processes begin in earnest. At early times ($t \lesssim 10-100\, \rm{Myr}$), the cluster loses mass and expands as a result of stellar winds and supernovae, reducing the potential of the cluster \citep[for further discussion, see][]{Weatherford2021}. As shown, higher metallicity models expand more significantly because of their enhanced mass loss. By $t \simeq 50-100\,\rm{Myr}$, black holes have mass-segregated to the center of their host cluster via dynamical friction. The timescale for this process scales inversely with the black hole mass, $t_{\rm seg} \simeq (\langle m \rangle /m_{\rm BH(Z)}) t_{\rm rel}$, where $\langle m \rangle\simeq 1~M_{\odot}$ is the mass of the average cluster member, $\langle m_{\rm BH(Z)}\rangle$ is the metallicity-dependent average black hole mass, and $t_{\rm rel}$ is the two-body relaxation time. Low-mass remnants experience weaker dynamical friction and sink to the cluster center more slowly than heavier black holes. Thus, the lowest-metallicity clusters begin black hole collapse episodes at earlier times relative to higher-metallicity clusters. Note that the black hole collapse described here refers specifically to the black hole sub-cluster \citep[e.g.,][]{Morscher2015} and is distinct from the core collapse of the cluster as a whole, which occurs on much longer timescales.
 
 On timescales of $1\,$Gyr and longer, black hole binary burning injects dynamical energy that heats the cluster and drives expansion of the core \citep[e.g.,][]{Mackey2008,BreenHeggie2013,Kremer2020_bhburning}. The larger and more massive black hole populations in low-metallicity clusters heat the cluster more significantly, producing greater core expansion. Interestingly, this metallicity dependent behavior in core radii is reflected in Milky Way globular clusters. The horizontal blue and red lines in this panel show the average core radii of Galactic clusters with masses in the range $[1-2]\times10^5\,M_{\odot}$ (similar to the masses of the eight \texttt{CMC} models shown) grouped into low and high metallicities with boundary at [Fe/H]$=-0.75$ \citep[cluster properties taken from][]{Harris1996}. We find average core radius values of $0.62 \pm 0.35\,$pc and $1.40\pm1.20\,$pc for the red and blue populations, respectively. Previous studies have also identified a similar trend for the globular clusters in M31 \citep{Barmby2002}. Granted, these observed values exhibit quite significant standard deviation, and of course other factors such as initial cluster concentration also contribute. Nonetheless, consistent with our models, low metallicity clusters appear to exhibit on average larger core radii compared to the high metallicity clusters; this may tentatively connect to the black hole-driven core expansion illustrated in our models. At later times ($t \gtrsim 10 \,\rm{Gyr}$), once the black hole populations have been depleted due to ejection, core radii begin to shrink as the clusters evolve toward global core collapse \citep[e.g.,][]{Kremer2019}. 

\begin{figure*}[ht!]
\centering
\includegraphics[width=0.9\linewidth,trim = 30 180 50 150,clip]{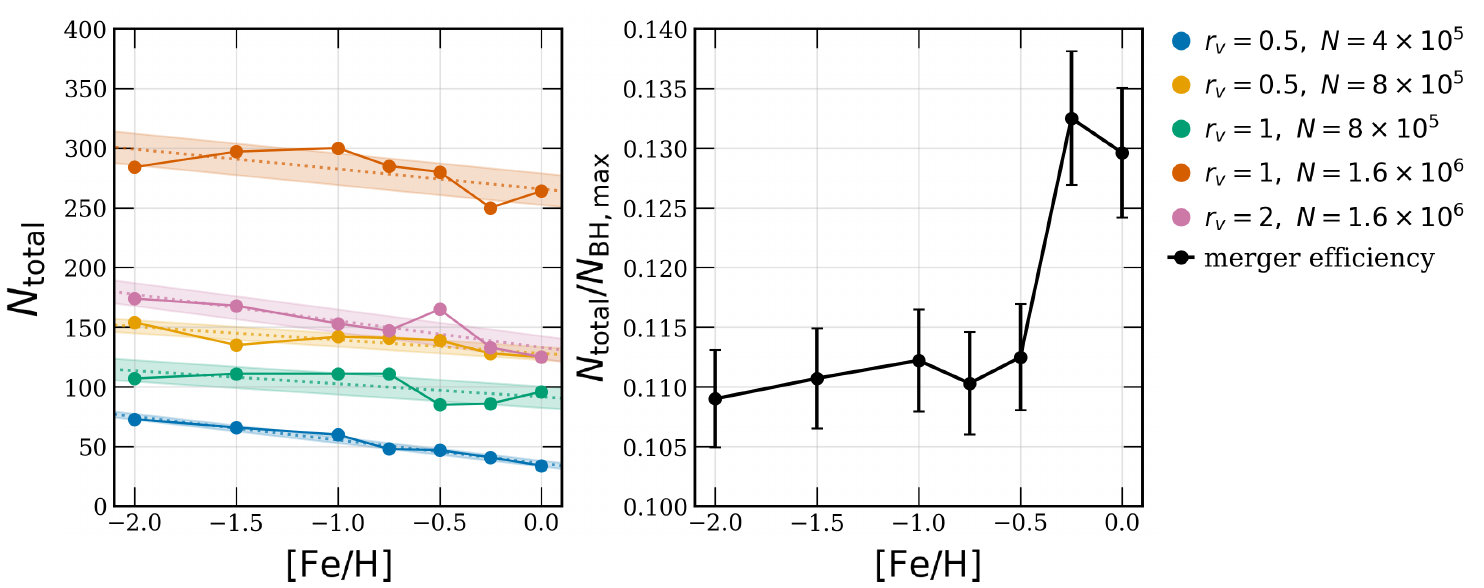}
\caption{Merger yield as a function of metallicity across 5 groups of \texttt{CMC} models with varying $N$ and $r_v$. \textit{Left panel:} Total number of black hole mergers (including linear fits in dashed lines and shaded 1$\sigma$ uncertainty). \textit{Right panel:} Total number of black hole mergers normalized by maximum number of black holes retained. Each point is summed over all models at a particular metallicity. Error bars represent Poisson uncertainty for each group of models. Although the total number of mergers declines with metallicity, the merger efficiency increases.}
\label{fig:merg_vs_met}
\end{figure*}

The bottom panel of Figure~\ref{fig:Nbh} shows the cumulative distribution of merger times for all dynamically-formed black hole mergers across the eight models. The shape of these distributions are distinct on both short and long timescales. On short timescales ($t\lesssim 100\,\rm{Myr}$), the merger times correspond to the mass segregation time of the black holes and formation of a central black hole subsystem. The time of the first merger in each model generally increases with metallicity; this reflects the delay in mass segregation time (see middle panel) associated with lower-mass black holes. On longer timescales ($t\gtrsim 1\,\rm{Gyr}$), the merger times are driven by the dynamical hardening timescale and the gravitational wave inspiral time of black hole binaries, which both decrease as black hole mass  increases \citep[e.g.,][]{Peters1964,Rodriguez2016}. As a result, the median merger time increases from $\sim 1\,\rm{Gyr}$ at low metallicities ([Fe/H]$\leq-1$) to $\sim 3\,\rm{Gyr}$ at solar metallicities ([Fe/H]$\geq0$). This shift in the delay-time distributions suggests that, for old globular clusters, high-metallicity systems similar to the observed ``red'' subpopulations (see Figure~\ref{fig:GC_FeH}), may contribute more mergers per cluster to the local binary black hole merger rate than lower-metallicity blue clusters \citep[see also][]{Chatterjee2017,Ye2026}.

\subsection{Effect of metallicity on black hole mergers}
\label{sec:bhmergers}

Metallicity also affects the dynamical pathways through which black holes form merging binaries. 
Figure~\ref{fig:metfit} shows the number of total and in-cluster binary black hole mergers as a function of initial cluster mass across four metallicities $[{\rm Fe/H}] = [0,\,-0.5,\,-1,\,-2]$. The models have varying $N=[2,\,4,\,8,\,16,\,32]\times10^{5}$ with otherwise identical initial conditions: $r_v = 1\,$pc and $R_{\rm gc} = 8\,$kpc (except for the highest mass clusters which all have $R_{\rm gc} = 20\,$kpc). In all cases, merger counts increase with cluster mass and are well-described by a power-law relation of the form $N_{\rm mergers} \propto M_{\rm cl}^{\gamma}$, with fitted power-law indices in the range $\gamma = 1.35-1.55$. We show the best-fit slope (identified via linear regression) for each metallicity group in each panel. The typical slope is consistent with the value of $1.4$ found in \citet{Mai2026} using \texttt{CMC} models, and other studies \citep{Hong2018,AntoniniGieles2020}. This scaling arises from two key effects: First, more massive clusters simply contain more stars and therefore form more black holes (this would imply a simple linear scaling with mass). Second, more massive clusters have deeper potential wells and therefore can more effectively retain black holes that are kicked via supernova physics, Newtonian dynamical interactions, and gravitational-wave recoil. These dynamical effects all contribute to the super-linear scaling of merger number with cluster mass.
Despite the strong influence of metallicity on the black hole mass function (see Figure~\ref{fig:bhmass}), the power-law scaling between cluster mass and merger productivity is largely insensitive to metallicity across the range explored here. This suggests that initial cluster mass is the dominant parameter governing total black hole merger output.

While the slope of the merger-mass relation remains roughly invariant with metallicity, the total number of mergers (intercept of the curves in Figure~\ref{fig:metfit}) does vary across metallicities. The left panel of Figure~\ref{fig:merg_vs_met} shows the total number of binary black hole mergers as a function of metallicity for five groups of our models with varying $[N,r_v]$ pairings : $N=[4,\,8,\,16]\times10^{5}$ and $r_v = [0.5,\,1,\,2]$ pc, all with $R_{\rm gc} = 8\,$kpc. All groups exhibit an overall decrease in the total number of mergers with increasing metallicity. This is primarily because high metallicity clusters form and retain smaller and less massive black hole populations (see top panel of Figure~\ref{fig:Nbh}).

Interestingly, although the total number of mergers decreases with metallicity, the merger efficiency --- defined as the number of mergers per retained black hole --- \textit{increases} with metallicity. This is illustrated in the right panel of Figure~\ref{fig:merg_vs_met}. Since the individual model merger fractions exhibit significant scatter, each scatter point has been computed by summing over all models at the same metallicity, with error bars showing the Poisson uncertainties. At low metallicity, the mass function of black holes broadens (see Figure~\ref{fig:bhmass}). In this regime, massive black holes mass segregate rapidly, and efficiently kick out a larger number of black holes en route to merger. The narrower (and lower-mass) black hole mass functions at higher metallicities are not as efficient at ejecting other black holes dynamically.\footnote{This is essentially a less extreme version of white dwarf merger dynamics in core-collapsed clusters \citep{Kremer2021_wd}.} Therefore, even though low-metallicity clusters form more black holes and more mergers in absolute terms, many black holes are removed before they can contribute to the total merger rate. In contrast, high-metallicity clusters form fewer and lower-mass black holes. These remnants receive weaker dynamical kicks from binary interactions and can be retained for longer times, increasing the merger efficiency when normalized by the total number of black holes retained.

\begin{figure*}[ht!]
\centering
\includegraphics[width=1\textwidth,trim = 17 110 10 80,clip]{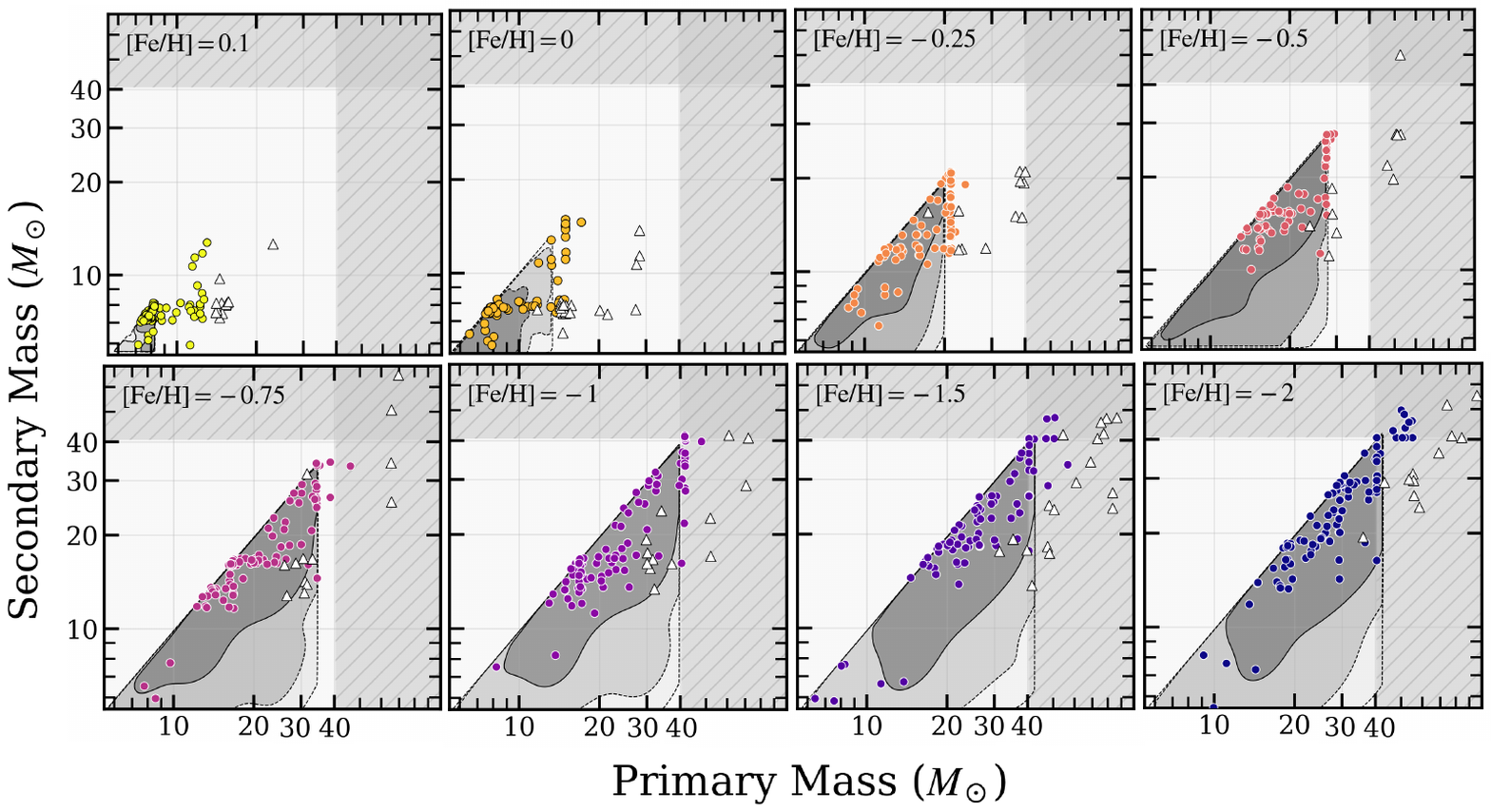}
\caption{Secondary versus primary mass for all black hole mergers across metallicities, again showing the same eight models as in Figure~\ref{fig:bhmass}. Filled circles represent 1G+1G mergers and empty triangles represent hierarchical mergers with at least one 2G+ component. Contours represent black hole binaries formed in isolated binary models run using \texttt{COSMIC} with the same stellar/binary evolution assumptions as our \texttt{CMC} models; the different contours show $1^{\rm st}$, $10^{\rm th}$, $50^{\rm th}$ percentiles.}
\label{fig:m1m2all}
\end{figure*}
 
Figure~\ref{fig:m1m2all} shows the secondary versus primary mass of all black hole mergers in the eight models studied in Section~\ref{sec:bhmass} at different metallicities, distinguishing between 1G+1G mergers (circles) and hierarchical mergers (triangles). Of the $111$ hierarchical mergers in these models, $106$ are 1G+2G mergers and only five are 2G+2G. Higher generation mergers become increasingly common in more massive clusters \citep[see Table~\ref{table:1} and discussion in][]{Mai2026}. Near solar metallicity (${ \rm [Fe/H]} = [0,0.1]$), mergers are tightly clustered at low masses ($m_1 \lesssim 20\,M_\odot$). As metallicity decreases, the merger population spreads out to higher masses, with masses ultimately extending to $m_1 =40.5\,M_\odot$, our assumed boundary for the pair-instability mass gap \citep[following][]{Belczynski2016}. This trend is qualitatively consistent with Figure~\ref{fig:bhmass}: the metallicity dependence that we see in the black hole mass function is inherited by the black hole merger population.

The gray contours show the distribution (1$^{\rm st}$, 10$^{\rm th}$, 50$^{\rm th}$ percentiles) of all binary black holes formed in our supplementary isolated binary evolution models using \texttt{COSMIC} (see Section~\ref{sec:methods}). 
\texttt{CMC} mergers extend to higher masses than isolated binaries, which arise for several reasons: (i) hierarchical mergers in clusters (triangles) provide a natural pathway for forming massive black hole binaries that would not form in isolation; (ii) at high metallicities, increased mass loss from stellar winds can unbind isolated systems, inhibiting formation of massive binary black holes in the isolated populations (this is most pronounced at [Fe/H]$\geq-0.25$ where the upper boundaries in primary mass for the $1^{\rm st}$ percentile \texttt{COSMIC} contours fall below the most massive 1G+1G \texttt{CMC} mergers); (iii) a small subset of 1G+1G mergers in each \texttt{CMC} population are formed via pre-collapse stellar collisions, enabled by the high densities of the cluster environments \citep[for further discussion, see][]{DiCarlo2019,Kremer2020_collisions,Gonzalez2021}; (iv) in a cluster, the most massive black holes preferentially segregate to their host's center, where they are more likely to dynamically assemble into similar-mass black hole binaries and merge. This fourth point also explains why mergers from clusters are more concentrated at mass ratios near unity compared to isolated binaries. 

\subsection{Comparison to GW241011 and GW241110}
\label{sec:comparison}

\begin{figure*}[ht!]
\centering
\includegraphics[width=\textwidth, trim = 5 250 50 92,clip]{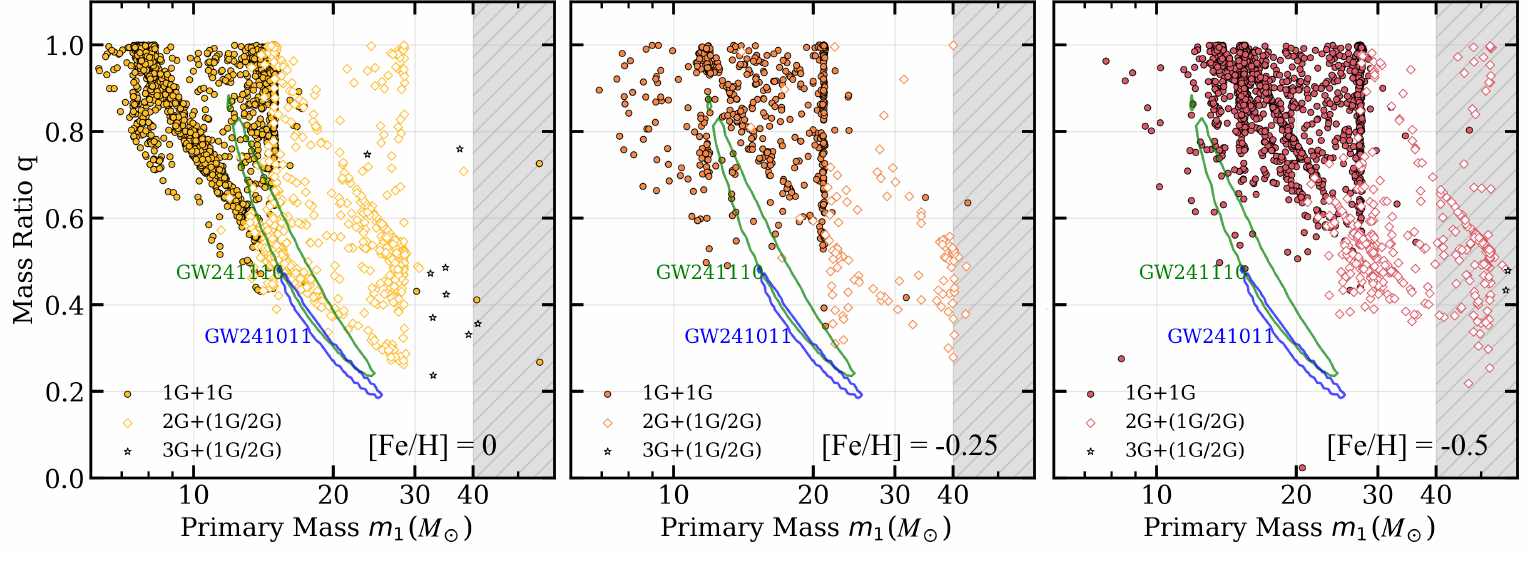}
\caption{Mass ratio $(q=m_2/m_1)$ versus primary mass $m_1$ for all black hole mergers in our highest-metallicity \texttt{CMC} models, from left to right [Fe/H]$=[0, -0.25,-0.5]$. Here we combine all models in Table~\ref{table:1} for these metallicity values. Blue and green contours show the $90$th percentile posteriors for the LVK events GW241011 and GW241110 respectively. Filled circles denote 1G+1G mergers, while diamonds (stars) denote mergers with at least one 2G (3G+) component.}
\label{fig:m1m23_q}
\end{figure*}

The two recent LVK events, GW241011 and GW241110, are notable for their unequal component masses, rapidly spinning primary black holes, and evidence for spin-orbit misalignment \citep{LIGO2025GW241011}. GW241011 contains one of the most rapidly spinning black holes observed to date with primary spin magnitude $\chi_1 = 0.78^{+0.09}_{-0.09}$, or equivalently $\chi_1>0.69$ at $95\%$ credibility, while GW241110 has a primary spin magnitude $\chi_1 = 0.61^{+0.33}_{-0.40}$ and shows evidence for a misaligned primary spin. These properties may point toward hierarchical origins because black hole merger remnants inherit angular momentum from the prior binary orbit and are expected to have dimensionless spins of order $\chi \sim 0.7$ \citep[e.g.,][]{GerosaFishbach2021}.

Figure~\ref{fig:m1m23_q} shows mass ratio, $q=m_2/m_1$, versus primary mass, $m_1$, for all mergers occurring in our highest metallicity \texttt{CMC} models; from left to right, [Fe/H]$=0$, $-0.25$, and $-0.5$. Here we show the black hole mergers occurring in all models shown in Table~\ref{table:1} (each panel combines multiple $N$ and $r_v$ values at each respective metallicity). 
The middle panel contains fewer mergers because our model set does not extend to as high of cluster mass at [Fe/H]$=-0.25$ (see Figure~\ref{fig:GC_FeH}). The left panel ([Fe/H]$=0$) also includes an extra massive model with initial $N=10^7$, increasing the total number of mergers, and especially the number of 3G+ mergers \citep[see][]{Mai2026}. 

Across all metallicities, the majority of first generation mergers fall within the mass ratio range $q \approx 0.6$--$1$, whereas second generation mergers peak at $q\sim 0.5$ and extend to smaller mass ratios. This behavior is expected from hierarchical mergers as the black hole remnant is generally more massive than the remaining black holes in the cluster, leading to mergers with more unequal mass binaries \citep[e.g.,][]{Rodriguez2019}. The contours show the $90^{\rm th}$ percentile posterior distributions for the two LVK events, GW241011 and GW241110 \citep{LIGO2025GW241011}. Both events lie at relatively low primary masses, but their unequal mass ratios place them in the region populated by hierarchical mergers. We note that GW241011 extends to lower mass ratios than are populated by our simulated mergers; this is likely due to assumptions made for the minimum mass of black holes in \texttt{COSMIC} rather than a true astrophysical lower limit. Both of the events fall within the locus of simulated mergers from high-metallicity clusters, with their low primary masses well reproduced by the second generation hierarchical mergers from our models. This is consistent with the analysis presented in \citet{LIGO2025GW241011} based on the original \texttt{CMC Cluster Catalog} models and reiterates the support for a high-metallicity cluster origin for both events.

\section{Summary and Discussion}
\label{sec:discussion}

\subsection{Summary}
Taken together, these results highlight several key trends concerning how metallicity impacts the black hole populations of dense stellar clusters:
\begin{enumerate}
    
    \item High-metallicity clusters produce fewer and lower-mass black holes because stars in high metallicity clusters lose more mass prior to collapse due to enhanced stellar wind mass loss, ultimately yielding smaller black holes. The resulting black hole \textit{merger} mass function is similarly shifted towards lower masses at higher metallicities.
    The black hole subsystems in low-metallicity clusters undergo mass segregation at earlier times and also feature shorter delay-time distributions, both due to the presence of more massive black holes. At later times, low-metallicity clusters also experience stronger core expansion as massive black hole binaries inject more energy through binary burning. 
    \item The slope of the relation between number of black hole mergers and initial cluster mass remains roughly constant ($N_{\rm merger} \propto M^{1.4})$ across different metallicities.
    For fixed cluster mass, we find that the total number of mergers per cluster decreases with metallicity, mainly because lower-metallicity clusters form and retain more black holes. However, when the merger yield is normalized by the number of retained black holes, we find the opposite trend: black holes in higher-metallicity clusters merge more efficiently, arising because lower-mass black hole binaries are less efficient at dynamically ejecting objects from their host cluster.
    \item High metallicity clusters produce low-mass binary black hole mergers with component masses and spins consistent with LVK events such as GW241011 and GW241110.
\end{enumerate}

\subsection{Discussion}

There are several important caveats to this analysis. First, the models in this grid assume a 5\% initial binary fraction across all masses, consistent with the original \texttt{CMC Cluster Catalog} models. The assumed binary fraction can affect the total black hole merger yield and the relative importance of different dynamical formation channels \citep[for recent studies of higher primordial binary fractions and their role in cluster evolution, see][]{GonzalezPrieto2024,Kiroglu2025a,Kiroglu2025,OConnor2026}. Therefore, the absolute number of black hole mergers and the relative contributions of different dynamical channels should be interpreted in the context of this assumed initial binary fraction. Second, our results, especially our precise black hole mass functions (see Figure~\ref{fig:bhmass}), depend crucially on the stellar and binary evolution prescriptions implemented in the version of \texttt{COSMIC} used in these simulations and we do acknowledge some assumptions are out-of-date \citep[for example, see more recent work in][]{GiacobboMapelli2018,Fragos2023,Andrews2025}.
 For example, LVK detections of objects in the lower black hole mass gap suggest fallback prescriptions similar to the ``delayed'' mechanism of \citet{Fryer2012} may be more appropriate than the ``rapid'' model used here. This would lead to formation of more low-mass black holes at all metallicities \cite[see, e.g.,][for discussion in the context of \texttt{CMC} models]{OConnor2026_LVK} and may help explain the slight discrepancy between our models and GW241011 shown in Figure~\ref{fig:m1m23_q}. 
That said, the general and key result that black hole masses decrease with increasing metallicity is expected to be robust, and is largely supported by the most up-to-date models and observations.

For this paper, we have focused on black holes and black hole mergers, but these models are in principle useful for studying a number of other cluster properties and compact-object sources. For example, significant populations of low mass X-ray binaries (LMXBs) are observed in Galactic and extragalactic globular clusters \citep[e.g.,][]{Kundu2002,Pooley2003,Heinke2003,Maccarone2007}.
\citet{Kundu2003} showed that metal-rich clusters are roughly ten times more likely to host bright LMXBs than metal-poor clusters, a trend not yet well understood from a theory perspective. These models can serve as a valuable tool for investigating this trend further. In addition, metallicity also impacts the populations of other compact binaries. Neutron star and white dwarf systems are robustly observed in their own right in globular clusters, and participate in dynamical processes including exchanges, binary hardening, and mergers analogous to black holes. We reserve for future work more detailed study of these lower-mass remnants and their associated observables.

\section{Data Availability}

The compiled catalog of all the binary black hole mergers formed in the models mentioned in this paper is available on \href{https://doi.org/10.5281/zenodo.20739155}{Zenodo} (DOI: 10.5281/zenodo.20739155). Each row in the catalog corresponds to one binary black hole merger and includes both the initial properties of the host cluster model and the properties of the merging binaries. The merger properties include merger time, component IDs, masses, spins, and generations of the two merging objects and their product. The catalog also contains kick velocity, cluster escape velocity at the time of merger, merger channel, and the final semi-major axis and eccentricity of the binary before merger.  

\acknowledgments
This research was supported in part by NSF grants PHY-2309135 to the Kavli Institute for Theoretical Physics (KITP), and AST-2511543 to Northwestern University.
F.K.\ and C.E.O.\ acknowledge support from CIERA Postdoctoral Fellowships.
M.Z.\ acknowledges funding from the Brinson Foundation in support of astrophysics research at the Adler Planetarium. 
C.P.\ is supported by an NSF Graduate Research Fellowship under grant DGE-2141064.
E.G.P.\ is supported by an  NSF Graduate Research Fellowship under grant DGE-2234667. C.S.Y.\ acknowledges support from the Alfred P. Sloan Foundation.
This work used computing resources at the San Diego Supercomputer Center, and at Northwestern's Quest high-performance computing facility, provided by CIERA under NSF grant PHY-2406802.

\appendix

\section{\texttt{CMC} model outputs}
\label{appA}
In Table~\ref{table:1} we list all initial properties and binary black hole merger information for all models computed in this study, as well as a subset of models from \citet{Kremer2020_catalog} for comparison.

\section{\texttt{COSMIC} initial conditions}
\label{appB}
Physical prescriptions and choices for various
parameterizations in the COSMIC models are summarized
below:

\textit{Initial conditions} --- We sample the primary mass of the binary with a Kroupa initial mass function~\citep{Kroupa2001} out to a maximum mass of $150~M_\odot$. Secondary masses are drawn from a uniform distribution in the mass ratio $q = m_2/m_1,\ m_2 \leq m_1$, with a minimum mass ratio that ensures the pre-main sequence timescale of the secondary does not exceed the (single-star) lifetime of the primary. Orbital periods and eccentricities follow \citealt{Sana2012}. We assume a value of $Z=0.02$ for Solar metallicity, and all systems are evolved for 13.7~Gyr.

\textit{Winds} --- We assume massive star metallicity-dependent line-driven winds from \citealt{Vink2001}, which include updated prescriptions for winds from luminous blue variable stars. Wolf-Rayet winds are assumed to follow \citealt{Vink2005}. Accretion of winds by the secondary are estimated assuming a Bondi-Hoyle mechanism~\citep{Bondi1944}, which depends on the wind velocity $v_\text{W}^2 = 2 \beta_\text{W} G M / R$ (where $G$ is the gravitational constant and $M$ and $R$ the stellar mass and radius). The constant factor $\beta_\text{W}$ can depend on stellar type, but we use a fixed value of $\beta_\text{W} = 0.125$ as in \citealt{Hurley2002}. 

\textit{Mass Transfer} --- Mass transfer stability is determined using stellar type-dependent critical mass ratios~\citep{Webbink1985}, for which we assume the values of \citealt{Hurley2002} with variations for (asymptotic) giant branch stars as in \citealt{Hjellming1987}. 
For stable mass transfer, the accretion rate onto the donor is limited by 10x the Kelvin-Helmholtz timescale~\citep{Hurley2002}. 
Mass lost during Roche-lobe overflow is assumed to carry away angular momentum from the system as if it were a wind from the donor, and accretion onto compact objects is Eddington-limited. 
For unstable mass transfer, we assume a standard $\alpha_\text{CE}-\lambda$ formalism with common-envelope efficiency $\alpha_\text{CE} = 1.0$ and a variable $\lambda$~\citep{Claeys2014}. 
An optimistic common envelope assumption is assumed, which allows Hertzsprung Gap stars to survive the common envelope phase~\citep{Belczynski2008}. 

\textit{Supernovae and Remnant Masses --- }
Supernova natal kicks are drawn from a Maxwellian distribution with a dispersion parameter of $265\,\text{km\,s}^{-1}$~\citep{Hobbs2005}, and are scaled down for black holes through fallback~\citep{Fryer2012}. 
Electron-capture supernovae are assumed to occur for helium core masses between $1.6-2.25 M_\odot$ at the end of helium burning, and kick magnitudes are drawn from a Maxwellian distribution with a dispersion parameter of $13.25\,\text{km\,s}^{-1}$ (i.e., 20x lower than standard core-collapse supernovae). 
Ultra-stripped supernovae are also accounted for if a helium star goes through a common envelope with a compact companion, with natal kick magnitudes drawn from the same distribution as electron-capture supernovae. 
The onset of pulsational pair instability supernovae is set to $45~M_\odot$; no black holes form above this limit in isolation. 
We assume a maximum neutron star mass (minimum black hole mass) of $2.5~M_\odot$, and determine remnant masses using the rapid explosion mechanism of \citealt{Fryer2012}. 
Neutrino mass loss in the supernovae follows \citealt{Zevin2020}. 

\textit{Population convergence} --- \texttt{COSMIC} continues sampling until pre-specified convergence criteria are reached for the target population \citep{Breivik2020}. In our models, the target population is binary black holes at their formation. We run models until the parameter space of primary mass--secondary mass--orbital period--eccentricity has stabilized such that additional draws from the population do not alter the shapes of these distributions significantly. 

\begin{deluxetable*}{ccccc|cc|ccccccc|cccc}

\tabletypesize{\scriptsize}
\setlength{\tabcolsep}{2pt}
\tablecaption{Summary of black hole formation and merger statistics across cluster models. The models indicated with an asterisk are models that were computed as part of the original \texttt{CMC Cluster Catalog}. For consistency, with \citet{Kremer2020_catalog,Mai2026}, we preserve the same model number labels used before this study: models 1--148 for the original \texttt{Catalog} and model 149 for \texttt{colossus} (the latter is indicated as $\dagger$). \label{table:1} }
\tablewidth{0pt}
\tablehead{
\colhead{model}&
\colhead{[Fe/H]} &
\colhead{$r_v$} &
\colhead{$R_{\rm gc}$} &
\colhead{$N$} &
\colhead{$N_{\rm BH}$} &
\colhead{$N_{\rm BH}$} &
\colhead{total} &
\colhead{primordial} &
\colhead{in-cluster} &
\colhead{2-body} &
\colhead{3-body} &
\colhead{4-body} &
\colhead{ejected} &
\colhead{1G+1G} &
\colhead{2G+1G} &
\colhead{2G+2G} &
\colhead{3G}\\
\\[-18pt]
\colhead{}&
\colhead{} &
\colhead{} &
\colhead{} &
\colhead{{$\times 10^{5}$}} &
\colhead{formed} &
\colhead{retained} &
\colhead{mergers} &
\colhead{mergers} &
\colhead{2-body} &
\colhead{capture} &
\colhead{capture} &
\colhead{capture} &
\colhead{mergers} &
\colhead{} &
\colhead{} &
\colhead{} &
\colhead{}
}
\startdata
 *14 & -2.0  & 0.5 & 8 & 4 & 754 & 543 & 65 & 0 & 14 & 0 & 5 & 11 & 35 & 55 & 10 & 0 & 0 \\
 150 & -1.5  & 0.5 & 8 & 4 & 739 & 510 & 66 & 0 & 21 & 0 & 1 & 12 & 32 & 51 & 13 & 2 & 0 \\
 *18 & -1.0  & 0.5 & 8 & 4 & 726 & 499 & 60 & 0 & 14 & 1 & 2 & 5  & 38 & 54 & 5  & 1 & 0 \\
 151 & -0.75 & 0.5 & 8 & 4 & 738 & 481 & 48 & 0 & 9  & 1 & 3 & 5  & 30 & 43 & 4  & 1 & 0 \\
 152 & -0.5  & 0.5 & 8 & 4 & 693 & 449 & 47 & 0 & 15 & 0 & 3 & 4  & 25 & 39 & 8  & 0 & 0 \\
 153 & -0.25 & 0.5 & 8 & 4 & 672 & 307 & 41 & 0 & 11 & 0 & 1 & 2  & 27 & 38 & 3  & 0 & 0 \\
 *22 &  0.0  & 0.5 & 8 & 4 & 589 & 326 & 34 & 0 & 6  & 0 & 1 & 1  & 26 & 31 & 3  & 0 & 0 \\
\hline
 *15 & -2.0  & 0.5 & 8 & 8 & 1574 & 1103 & 154 & 0 & 49 & 0 & 10 & 15 & 80 & 132 & 21 & 1 & 0 \\
 154 & -1.5  & 0.5 & 8 & 8 & 1551 & 1085 & 135 & 0 & 42 & 4 & 7  & 14 & 68 & 114 & 20 & 1 & 0 \\
 *19 & -1.0  & 0.5 & 8 & 8 & 1425 & 1023 & 141 & 0 & 41 & 1 & 8  & 9  & 82 & 119 & 21 & 1 & 0 \\
 155 & -0.75 & 0.5 & 8 & 8 & 1461 & 1018 & 141 & 0 & 37 & 2 & 11 & 12 & 79 & 119 & 21 & 1 & 0 \\
 156 & -0.5  & 0.5 & 8 & 8 & 1404 & 948  & 139 & 0 & 42 & 2 & 10 & 9  & 76 & 118 & 19 & 1 & 0 \\
 157 & -0.25 & 0.5 & 8 & 8 & 1368 & 768  & 128 & 0 & 35 & 5 & 10 & 8  & 70 & 107 & 19 & 2 & 0 \\
 *23 & 0.0   & 0.5 & 8 & 8 & 1238 & 776  & 125 & 0 & 59 & 2 & 2  & 5  & 57 & 101 & 22 & 1 & 0 \\
\hline
 *51 & -2.0  & 1.0 & 8 & 8 & 1622 & 1093 & 107 & 6 & 30  & 0  & 4  & 11 & 56  & 95  & 12  & 0 & 0 \\
 158 & -1.5  & 1.0 & 8 & 8 & 1589 & 1048 & 110 & 9 & 30  & 1  & 5  & 11 & 54  & 92  & 17  & 1 & 0 \\
 *55 & -1.0  & 1.0 & 8 & 8 & 1534 & 1035 & 111 & 9 & 30  & 3  & 6  & 6  & 57  & 98  & 13  & 0 & 0 \\
 159 & -0.75 & 1.0 & 8 & 8 & 1517 & 1008 & 111 & 4 & 30  & 1  & 7  & 9  & 60  & 98  & 11  & 2 & 0 \\
 160 & -0.5  & 1.0 & 8 & 8 & 1490 & 951  & 85  & 0 & 25  & 2  & 6  & 12 & 40  & 72  & 12  & 1 & 0 \\
 161 & -0.25 & 1.0 & 8 & 8 & 1452 & 697  & 86  & 1 & 24  & 1  & 12 & 4  & 44  & 75  & 10  & 1 & 0 \\
 *59 & 0.0   & 1.0 & 8 & 8 & 1289 & 781  & 95  & 0 & 44  & 3  & 4  & 3  & 41  & 76  & 19  & 0 & 0 \\
 174 & 0.1   & 1.0 & 8 & 8 & 1076 & 553  & 82  & 2 & 26  & 1  & 6  & 5  & 42  & 70  & 12  & 0 & 0 \\
\hline
 *52 & -2.0  & 1.0 & 8 & 16 & 3234 & 2305 & 284 & 6  & 85  & 14 & 20 & 21 & 138 & 238 & 45 & 1 & 0 \\
 162 & -1.5  & 1.0 & 8 & 16 & 3146 & 2233 & 297 & 12 & 115 & 10 & 19 & 16 & 125 & 247 & 47 & 3 & 0 \\
 *56 & -1.0  & 1.0 & 8 & 16 & 3064 & 2190 & 300 & 9  & 108 & 7  & 15 & 24 & 137 & 245 & 50 & 5 & 0 \\
 163 & -0.75 & 1.0 & 8 & 16 & 3029 & 2127 & 285 & 4  & 99  & 5  & 15 & 21 & 141 & 229 & 53 & 3 & 0 \\
 164 & -0.5  & 1.0 & 8 & 16 & 2984 & 2075 & 280 & 2  & 96  & 8  & 23 & 12 & 139 & 224 & 55 & 1 & 0 \\
 165 & -0.25 & 1.0 & 8 & 16 & 2935 & 1641 & 249 & 0  & 90  & 12 & 12 & 9  & 126 & 203 & 45 & 1 & 0 \\
 *60 & 0.0   & 1.0 & 8 & 16 & 2539 & 1623 & 264 & 2  & 118 & 4  & 13 & 13 & 114 & 208 & 50 & 5 & 1 \\
\hline
 *88 & -2.0  & 2.0 & 8 & 16 & 3200 & 2222 & 174 & 14 & 46 & 5  & 13 & 7  & 89  & 155  & 19 & 0 & 0 \\
 166 & -1.5  & 2.0 & 8 & 16 & 3140 & 2142 & 168 & 15 & 37 & 4  & 11 & 15 & 86  & 142  & 26 & 0 & 0 \\
 *92 & -1.0  & 2.0 & 8 & 16 & 3058 & 2079 & 153 & 9 & 45  & 5  & 10 & 9  & 75  & 138  & 13 & 2 & 0 \\
 167 & -0.75 & 2.0 & 8 & 16 & 2999 & 2003 & 147 & 5 & 40  & 2  & 9  & 13 & 78  & 128  & 19 & 0 & 0 \\
 168 & -0.5  & 2.0 & 8 & 16 & 2970 & 1942 & 165 & 4 & 54  & 3  & 16 & 6  & 82  & 138  & 27 & 0 & 0 \\
 169 & -0.25 & 2.0 & 8 & 16 & 2883 & 1402 & 133 & 9 & 46  & 3  & 10 & 8  & 57  & 107  & 25 & 1 & 0 \\
*96 & 0.0   & 2.0 & 8 & 16 & 2551 & 1463 & 125  & 5 & 35  & 3  & 7  & 6  & 69  & 105  & 18 & 2 & 0 \\
\hline
 *145 & -2.0  & 1.0 & 20 & 32 & 6427 & 4786 & 750 & 21 & 312 & 34 & 74 & 21 & 288 & 555 & 172 & 18 & 5 \\
 170  & -1.0  & 1.0 & 20 & 32 & 6093 & 4583 & 696 & 15 & 286 & 38 & 58 & 30 & 269 & 509 & 166 & 19 & 2 \\
 171  & -0.5  & 1.0 & 20 & 32 & 5927 & 4392 & 744 & 1  & 330 & 56 & 71 & 25 & 261 & 533 & 186 & 23 & 2 \\
 *147 &  0.0  & 1.0 & 20 & 32 & 5037 & 3692 & 572 & 4  & 280 & 32 & 31 & 12 & 213 & 408 & 141 & 22 & 1 \\
\hline
 *146 & -2.0 & 2.0 & 20 & 32 & 6337 & 4482 & 380 & 16 & 129 & 11 & 35 & 9 & 180 & 305 & 69 & 6  & 0 \\
 172 & -1.0  & 2.0 & 20 & 32 & 6066 & 4275 & 427 & 10 & 170 & 10 & 38 & 21 & 178 & 335 & 87 & 4  & 1 \\
 173 & -0.5  & 2.0 & 20 & 32 & 5888 & 4028 & 424 & 4  & 160 & 23 & 37 & 14 & 186 & 332 & 82 & 10 & 0 \\
 *148 & 0.0  & 2.0 & 20 & 32 & 5090 & 3191 & 364 & 8  & 169 & 12 & 32 &  8 & 135 & 278 & 77 & 8  & 1 \\
\hline
 175 & -0.5 & 1.0 & 8  & 2 & 358 & 211 & 11 & 0  & 3   & 0   & 1  & 2 & 5 & 11 & 0  & 0 & 0\\
 176 & -0.5 & 1.0 & 8  & 4 & 751 & 453 & 37 & 0  & 14   & 1   & 0  & 4 & 18 & 33 & 4  & 0 & 0\\
 \hline
 $^\dagger$149 & -1.0  & 2.0 & 20 & 100 & 18693 & 14475 & 1367 & 32 & 686 & 152 & 148 & 34 & 315 & 914 & 374 &  63 & 14\\
 177 & 0.0  & 2.0 & 20 & 100 & 15425 & 11922 & 1126 & 33 & 695 & 134 & 85 & 19 & 160 & 731 & 337 & 49 & 9\\
 \hline
\enddata
\end{deluxetable*}

\bibliographystyle{aasjournal}
\bibliography{mybib}

\end{document}